\documentstyle[aps,preprint]{revtex}
\begin{document}
\draft

\title{Corrections to Scaling in the Phase-Ordering Dynamics \\
       of a Vector Order Parameter}
\author{N. P. Rapapa and A. J. Bray}
\address{Department of Physics and Astronomy, The University, 
Manchester M13 9PL, UK}
\date{\today}
\maketitle

\begin{abstract}
Corrections to scaling, associated with deviations of the order parameter 
from the scaling morphology in the initial state, are studied for systems 
with $O(n)$ symmetry at zero temperature in phase-ordering kinetics. 
Including corrections to scaling, the equal time pair correlation function
has the form $C(r,t) = f_0(r/L) + L^{-\omega} f_1(r/L) + \cdots$, 
where $L$ is the coarsening length scale. The correction-to-scaling 
exponent $\omega$ and the correction-to-scaling function $f_1(x)$ are 
calculated for both nonconserved and conserved order parameter systems using 
the approximate Gaussian closure theory of Mazenko. In general 
$\omega$ is a non-trivial exponent which depends on both the dimensionality,  
$d$, of the system and the number of components, $n$, of the order parameter. 
Corrections to scaling are also calculated for the nonconserved 1-$d$ XY 
model, where an exact solution is possible.
\end{abstract}

\narrowtext


\section{Introduction}
The dynamics  of a system undergoing phase-ordering following a quench 
from the high temperature (disordered) phase to the ordered phase is of great 
interest\cite{Review}. The kinetics of systems with $O(n)$ symmetry 
subject to `Model A' dynamics \cite{HH}(i.e.\ systems with nonconserved order 
parameter) and `Model B' dynamics \cite{HH} (systems with conserved order 
parameter) have been previously studied \cite{BH1,FA} within a Gaussian closure 
theory originally developed by Mazenko \cite{MAZ1,MAZ2} following the 
seminal work of Ohta, Jasnow and Kawasaki (OJK) \cite{OJK}. In previous work 
\cite{BRC} we have computed the form of the corrections to the scaling limit, 
and the correction-to-scaling exponent, for a number of systems with 
nonconserved order parameter. These include some exactly soluble models, and 
the Model A dynamics of a scalar field within the Mazenko theory.

In the present work we turn our attention to systems with  a vector order 
parameter, both nonconserved and conserved. The corrections to scaling for 
systems with continuous symmetry will be calculated using the Mazenko theory. 
It should be mentioned that this approach has been shown to be more successful, 
at a quantitative level, in systems with nonconserved order parameter than 
those with conserved order parameter \cite{RPB}. Nevertheless, the results 
obtained in the conserved case are in qualitative agreement with those obtained 
in simulations. Furthermore, the Mazenko approach seems the only available 
method to probe the questions of corrections to scaling addressed here. 
In particular, we found that, for nonconserved scalar fields, the 
correction-to-scaling exponent $\omega$ is predicted by this approach to have 
a nontrivial value. We will show that this same feature is present for the 
vector fields, with and without conservation.

It is well established \cite{Review} that at late times most phase-ordering 
systems approach a scaling regime, where the equal-time pair correlation 
function $C(r,t) \equiv \langle\vec{\phi}({\bf x}+{\bf r},t)\cdot
\vec{\phi}({\bf x},t)\rangle$ takes the form $C(r,t)= f[r/L(t)]$. 
The characteristic length scale $L(t)$ grows with time as 
$L(t) \sim t^a$, where $a$ is the growth exponent which depends on the 
nature of the dynamics and the symmetry of the order parameter. 
In particular, $a = 1/2$ for nonconserved order parameter systems, while 
$a = 1/3$ for systems with conserved scalar order parameter and $a = 1/4$ 
for systems with conserved vector order parameter (with logarithmic 
corrections for $n = 2$, $d > 2$ \cite{BR}). In previous work \cite{BRC} 
we studied how scaling is approached in nonconserved order 
parameter models such as the 1-$d$ Ising model with Glauber dynamics, the 
$n-$vector model with  $n = \infty$, the approximate OJK theory and the 
Mazenko theory for scalar fields. In all these cases $\omega$ was found to 
be trivial ($\omega = 4$) except the last, for which $\omega$ was found to be 
non-trivial and dimensionality dependent. 
The relevance of corrections to scaling lies in interpreting 
experimental and simulation results, where it is advantageous to know how 
the scaling limit is approached. Corrections to scaling in systems with finite 
$n>1$  in $d =3$ and $d = 2$ were not considered in \cite{BRC}. 
The main objective in this article is to study systems with $n \ge 2$.

This article is devoted to the study of the corrections to scaling 
for systems with $O(n)$ symmetry in phase-ordering dynamics. The leading 
corrections to scaling enter the correlation function in the form 
\begin{equation}
C(r,t) = f_0(r/L)+ L^{-\omega}f_1(r/L), 
\end{equation}
where $f_0(x)$ is the `scaling function' and $f_1(x)$ the 
`correction-to-scaling function'. The quantity which unites theory, computer 
simulation and experiment is the structure factor 
$S(k,t) = L^d g_{0}(y) + L^{d - \omega} g_{1}(y)$, 
where $g_{0}(y)$ and $g_{1}(y)$ are the $d$-dimensional Fourier transforms of 
$f_{0}(x)$ and $f_{1}(x)$ respectively, and $y=kL$. Coniglio and Zannetti
\cite{CZ} solved the conserved $O(n)$ model for $n = \infty$ exactly, and 
found that no simple scaling exists. Instead a `multiscaling' behavior was 
obtained, raising the question of whether simple scaling exists in conserved 
order parameter systems with $n > 1$ generally (or even for conserved scalar 
fields). However, it was later shown by Bray and Humayun \cite{BH2}, 
analytically within the Mazenko theory, that scaling does exist for large 
but finite $n$. Attempts to find multiscaling behaviour in simulation date 
for conserved scalar fields \cite{CZ2}, or the conserved XY model in two 
\cite{MG} or three \cite{SR} dimensions were not successful. It is now 
generally believed that scaling is recovered asymptotically in time in the conserved 
$O(n)$ model, for all finite $n$, though multiscaling may be observable in 
the preasymptotic regime \cite{Zannetti}.

There are other sources of corrections to scaling apart from the one 
considered in this paper. In phase-ordering systems there is, in addition 
to the time-dependent coarsening scale $L(t)$, a second  
characteristic length  scale -- the `defect core size' $\xi$ -- in systems 
with topological defects. The corrections to scaling associated with nonzero 
defect core size (where $\xi$ is the domain wall thickness in scalar systems) 
are expected to enter as a power of ${\xi}/L$. Here we are interested 
primarily in the corrections to scaling associated with non-scaling initial 
conditions. We therefore suppress the contributions associated with nonzero 
core size $\xi$ by taking the `hard-spin' limit, i.e.\ working with an 
order-parameter field whose length is everywhere unity, $\vec{\phi}^2=1$, 
which forces $\xi=0$ (though in the Mazenko theory this limit will be taken 
at the end). Also thermal fluctuations at $T > 0$ may give rise to 
significant corrections to scaling for systems quenched to a nonzero final 
temperature $T$ \cite{Bray89} (where $0 < T < T_{c}$, with $T_{c}$ the critical 
temperature) as has been shown explicitly in the nonconserved $O(n)$ model with 
$n \to \infty$ \cite{BN}. However, we will only be studying systems quenched 
to $T = 0$. Although corrections to scaling due to thermal fluctuations and 
nonzero $\xi$ are important we will not consider them further in this paper.

The outline of the paper is as follows. In the following section 
the approximate Mazenko theory is discussed and some general concepts are 
introduced. Section III deals with nonconserved order parameter systems. 
In section IV, corrections to scaling for the nonconserved 1-$d$ XY model 
will be studied. Systems with conserved order parameter are considered in 
section V. Section VI concludes with a summary and discussion.

\section{Mazenko Theory}
A `Gaussian closure' theory, building on the earlier work of Ohta, Jasnow 
and Kawasaki \cite{OJK} has been developed by Mazenko \cite{MAZ1}. This theory 
has been successfully applied to $O(n)$ models in the theory of phase-ordering 
dynamics \cite{BH1,RPB}. The equation of motion for an order parameter $\vec\phi$ 
with continuous symmetry, for systems quenched to $T = 0$, is 
\begin{equation}
\frac {\partial {{\vec \phi}(1)}}{\partial t_{1}} =
(-\nabla^2_1)^p \left[\nabla^2_1{\vec \phi}(1)-
\frac{\partial V[{\vec \phi}(1)]}{\partial {\vec \phi}(1)} \right],  
\label{OPE}
\end{equation}
where $p=1$ and $p=0$ for conserved order parameter (Model B) and 
nonconserved order parameter (Model A) systems respectively. In (\ref{OPE}), 
$V({\vec \phi})$ is a symmetric double-well potential for the scalar case,  
and a `wine bottle' potential with a degenerate continuum manifold for a vector 
order parameter. Compact notation has been used in which 
`1' represents the space-time point $({\vec x}_1,t_1)$ and $\nabla^2_1$ 
means the Laplacian with respect to ${\vec x}_1$. Multiplying (\ref{OPE}) 
by ${\vec \phi}(2)$, averaging over initial conditions, and using the 
translational invariance of $C(12)$ gives (for $t_1 = t_2 =t)$
\begin{equation}
\frac{1}{2}\,\frac {\partial {C(12)}}{\partial t} =
(-{\nabla^2})^p \left[\nabla^2 C(12)-
\langle \frac{\partial V[\vec\phi(1)]}{\partial\vec\phi(1)} 
\cdot \vec\phi(2) \rangle \right],  
\label{CFE}
\end{equation}
where now $\nabla^2$ is the Laplacian with respect to 
$r = |{\bf x}_{1} -{\bf x}_{2}|$ and $C(12) = 
\langle {\vec \phi}(1)\cdot{\vec \phi}(2) \rangle $. 
The angular brackets denote the average over the initial conditions. 
In order to evaluate the average of the last term in (\ref{CFE}) one 
introduces an auxiliary field ${\vec m}(r,t)$  related to 
$\vec \phi$ by $\nabla^2_m \vec \phi = 
2\,\partial V(\vec \phi)/\partial\vec\phi$, with boundary 
condition $ {\vec \phi} \to \vec{m}/|\vec{m}|$ as $|\vec m| \rightarrow 
\infty$, and $\vec\phi=0$ at $\vec{m}=0$. Near a defect, the field 
$\vec{m}({\bf r})$ is the position vector of the point ${\bf r}$ in 
the plane normal to the defect. The assumption that ${\vec m}$  is a 
Gaussian field enables the evaluation of the average of the last term 
on the right hand side of (\ref{CFE}) giving \cite{BH1}
\begin{equation}
\frac{1}{2}\,\frac {\partial {C(12)}}{\partial t} =
(-\nabla^2)^p \left[\nabla^2 C(12)+
 \frac{1}{2S_{0}(1)}\,\gamma\,\frac{dC(12)}{d\gamma}\right],  
\label{ACFE}
\end{equation} 
where $S_{0} = \langle m(1)^2 \rangle$ and ${\gamma}(12) = 
\langle m(1)m(2) \rangle /[\langle m(1)^2 \rangle \langle m(2)^2 
\rangle]^{1/2}$ is the normalised correlator of the field $m$ (where $m$ 
is one of the components of ${\vec m}$). An explicit 
expression which relates $\gamma$ to $C(12)$ was given in \cite{BPTO}
\begin{equation}
C = \frac{n{\gamma}}{2{\pi}}\,
\left[B\left(\frac{n+1}{2},\frac{1}{2}\right)\right]^2\,
F\left(\frac{1}{2},\frac{1}{2};\frac{n+2}{2};{\gamma}^2\right),
\label{BPT}
\end{equation} 
where $B(y,z) = \Gamma(y)\Gamma(z)/\Gamma(y+z)$ is the Beta function 
and $F(a,b;c;z)$ the hypergeometric function. Equations (\ref{ACFE}) and 
(\ref{BPT}) provide closed form equations for $C(12)$. On substituting 
(\ref{BPT}) in (\ref{ACFE}) one obtains an equation for $\gamma$ which can 
in principle be solved numerically and substituted back into  (\ref{BPT}) 
to obtain the correlation function $C(12)$. We note at this point that 
in deriving the correlation function (\ref{BPT}), the `hard-spin' limit 
$\phi = \vec{m}/|\vec{m}|$ was employed. Since this result holds 
far from defect cores, it will correctly describe the scaling limit where 
the defects are dilute. Here we are also using it to compute the corrections 
to scaling.

\section{Nonconserved O(n) Model}
For a nonconserved system $p = 0$, and equation (\ref{ACFE}) is simply
\begin{equation}
\frac{1}{2}\,\frac {\partial {C(12)}}{\partial t} =
{\nabla^2}C(12)+ \frac{1}{2S_{0}(1)}\,\gamma\,\frac{dC(12)}{d\gamma}  
\label{CCCE}
\end{equation} 
For $n = 1$, using the properties of the hypergeometric function the last 
term on the right hand side of (\ref{CCCE}) can be written in terms of 
$C(12)$ only, resulting in an equation which is independent of ${\gamma}(12)$. 
Corrections to scaling in this case where obtained in our previous work 
\cite{BRC}, and will not be considered further here. For general $n$, 
$\gamma$ cannot be eliminated in favour of $C(12)$, and we will 
therefore work with $\gamma$ instead of $C(12)$. From dimensional 
considerations we see that $S_{0} \sim L^2$ and can be chosen as $S_{0} = 
{L^2}/{\lambda}$. This choice effectively defines $L$, up to an overall 
constant. For $n \rightarrow \infty$, an expansion in $1/n$ can be 
performed on $C(\gamma)$, and in this limit $\gamma\,dC/d\gamma = 
C + {C^3}/n + O(1/n^2)$. For $n = \infty$, Mazenko theory reduces to the 
$n = \infty$ $n$-vector model for which an exact solution, including the 
corrections to scaling, is known \cite{BRC}. Expressing (\ref{CCCE}) in 
terms of ${\gamma}$ explicitly leads to 
\begin{equation}
\frac{1}{2}\,\frac {\partial {\gamma}}{\partial t} = 
\frac{C_{\gamma \gamma}}{C_{\gamma}}\,\left(\frac{\partial \gamma}
{\partial r}\right)^2 + \frac{\partial^2 \gamma}{\partial r^2}
+\frac{d-1}{r}\,\frac {\partial {\gamma}}{\partial r}+
\frac{\lambda}{2L^2}\,\gamma\ ,  
\label{CCE}
\end{equation} 
where $C_{\gamma} = dC/{d{\gamma}}$ etc. Since $C(r,t)$ is a function of 
$\gamma(r,t)$, the scaling and corrections to scaling can be imposed on 
$\gamma(r,t)$. In the scaling limit we expect $\gamma(r,t)$ to approach 
the scaling function ${\gamma_{0}}(r/L)$ which is $L$-independent if all 
lengths are scaled by $L$. In this limit therefore one expects 
$LdL/dt = $ constant. Including corrections to scaling in $\gamma(r,t)$ 
and $L(t)$ as usual \cite{BRC} we can write
\begin{eqnarray}
\gamma(r,t) & = & \gamma_0\left(\frac{r}{L}\right) + L^{-\omega}\,
\gamma_1\left(\frac{r}{L}\right) + \cdots, \\
C(r,t) & = & f_0\left(\frac{r}{L}\right) 
+ L^{-\omega}\,f_1\left(\frac{r}{L}\right) + \cdots, \\
\frac{dL}{dt} & = & \frac{1}{2L} + \frac{b}{L^{1+\omega}} + \cdots, 
\label{L}
\end{eqnarray}
where 
\begin{eqnarray}
\label{S}
f_0\left(\frac{r}{L}\right) &=& C(\gamma_0), \\
f_1\left(\frac{r}{L}\right) &=& {\gamma_{1}}\left(\frac{r}{L}\right)\,
{\left[\frac{dC}{d\gamma}\right]}_{\gamma = \gamma_0},
\label{CS}
\end{eqnarray}
and $b$ is a constant. Equating leading and next-to-leading powers of $L$ 
in the usual way gives 
\begin{equation}
{\gamma_0}'' + \frac{C_{\gamma_{0} \gamma_{0}}}{C_{\gamma_{0}}}\,
{{\gamma_{0}}'}^2 + \left[\frac{x}{4}+\frac{d-1}{x} \right ]\,{{\gamma_{0}}'}
+\frac{\lambda}{2}\,\gamma_{0}=0,
\label{NSE}
\end{equation} 
 \begin{eqnarray}
{\gamma_1}'' +  \left[\frac{x}{4}+\frac{d-1}{x} \right ]\,
{{\gamma_{1}}'}+\left [\frac{\lambda}{2}+\frac{\omega}{4} \right]\, 
\gamma_{1} + \frac{b}{2}\,x\,{{\gamma_{0}}'} +     \nonumber\\
2\,\frac{C_{\gamma_{0} \gamma_{0}}}{C_{\gamma_{0}}}\,
{{\gamma_{0}}'}\,{{\gamma_1}'}
+ \left [ \frac{C_{\gamma_{0} \gamma_{0} \gamma_{0}}}{C_{\gamma_{0}}} 
 - \frac{(C_{\gamma_{0} \gamma_{0}})^2}
{{C_{\gamma_0}}^2} \right]\,\gamma_{1}\,{{\gamma_0}'}^2=0, 
\label{NCSE}
\end{eqnarray} 
with $C_{\gamma_{0}} = [dC/d\gamma]_{\gamma=\gamma_0}$ etc. The primes 
indicate derivatives with respect to the scaling variable $x = r/L$.

Equations (\ref{NSE}) and (\ref{NCSE}) are to be integrated numerically 
subject to appropriate `initial' conditions imposed at $x=0$. Since $x=0$ 
corresponds to $\gamma_0=1$, the initial conditions are obtained by 
considering the regime ${\gamma_0} \rightarrow 1$. Using the properties 
of the hypergeometric functions\cite{ABS} one can derive relations between 
$C({\gamma_0})$ and its derivatives as ${\gamma_0} \rightarrow 1$. 
Up to prefactors of order unity, we find in this limit 
\begin{eqnarray}
C_{\gamma_0\gamma_0}/C_{\gamma_0} & \sim & [(1-\gamma_0)
|\ln (1-\gamma_0)|]^{-1},\ \ \ n=2 \nonumber \\
C_{\gamma_0\gamma_0\gamma_0}/C_{\gamma_0} & \sim & 
[(1-\gamma_0)^2|\ln(1-\gamma_0)|]^{-1}\, \ \ \  n = 2 \nonumber\\
C_{\gamma_0\gamma_0}/C_{\gamma_0} & \sim & [1-\gamma_0]^{(n-4)/2}, 
\ \ \ 2<n<4 \nonumber \\
C_{\gamma_0\gamma_0\gamma_0}/C_{\gamma_0} &\sim &
[1-\gamma_0]^{(n-6)/2},\ \ \  2 < n < 4 \nonumber\\
C_{\gamma_0\gamma_0}/C_{\gamma_0} &\sim& 
|\ln(1-\gamma_0)|, \ \ \ n=4 \nonumber\\
C_{\gamma_0\gamma_0\gamma_0}/C_{\gamma_0}&\sim & [1-\gamma_0]^{-1},
\ \ \ n = 4 \nonumber\\
C_{\gamma_0\gamma_0}/C_{\gamma_0} & \to & {\rm constant},
\ \ \ 4 < n < 6 \nonumber\\
C_{\gamma_0\gamma_0\gamma_0}/C_{\gamma_0}&\sim & 
[1-\gamma_0]^{(n-6)/2},\ \ \ 4 < n < 6, 
\label{SMLC}
\end{eqnarray} 
and so on. We have given explicit  expressions for $C_{\gamma_0\gamma_0} 
/C_{\gamma_0}$ and $C_{\gamma_0\gamma_0\gamma_0} /C_{\gamma_0}$ as 
$\gamma_0 \rightarrow 1$ for the values of $n$ which we are going to study. 
Using the above results one can show \cite{BH1} that the small-$x$ behavior 
of $\gamma_0(x)$ is given by 
\begin{equation}
\gamma_{0}(x) = 1- \frac{\lambda}{4d}\,x^2 + \cdots 
\end{equation}
for $n \ge 2$,  where the limiting forms in (\ref{SMLC}) were used to 
demonstrate that the term involving $C_{\gamma_0\gamma_0}/C_{\gamma_0}$ 
in (\ref{NSE}) is subdominant as $x \to 0$ for $n \ge 2$.

For large-$x$, $\gamma_{0} \rightarrow 0$ (also $C(12) \rightarrow 0$) and 
equation (\ref{NSE}) becomes linear because in this limit the second term 
in (\ref{NSE}) is negligible. It is easy to show that two linearly  
independent solutions of the linearised equation have the asymptotic 
forms $\gamma_{01} \sim x^{-2\lambda}$ and 
$\gamma_{02} \sim x^{2\lambda - d}\exp(-x^2/8)$, for $x \to \infty$. 
As equation (\ref{NSE}) is integrated forward from $x = 0$, the large-$x$ 
solution obtained will in general be a linear combination of $\gamma_{01}$ 
and $\gamma_{02}$. The amplitudes of $\gamma_{01}$ and $\gamma_{02}$, however, 
depend on $\lambda$. For systems with initial conditions containing only 
short-range spatial correlations (as is the case for systems quenched from 
high temperature), a power-law decay is unphysical, and $\lambda$ is 
determined by the condition that the coefficient of the power-law term, 
$\gamma_{01}$, must vanish \cite{BH1}. Note that $\lambda$ is related to the 
exponent $\bar{\lambda}$ describing the decay of the autocorrelation function 
\cite{LIMAZ} via ${\bar\lambda} = d - \lambda$. Values for $\lambda$ are 
given in Table 1 for $2 \le n \le 5$ and $1 \le d \le 3$. Comparison 
of the predicted values of $\lambda$ with simulations \cite{NBMH} and 
experiments \cite{MPY} show reasonable agreement. It can be shown 
that for $d \rightarrow \infty$ the OJK result ($\lambda = d/2$) is 
recovered for both scalar \cite{LiuMaz} and vector \cite{BH1} cases. 
The same limit for $\lambda$ is also obtained for $n  \rightarrow \infty$ 
at arbitrary $d$. 

The correction-to-scaling exponent, $\omega$, is found from (\ref{NCSE}) 
in a similar way to the determination of $\lambda$ from (\ref{NSE}). 
In order to specify initial conditions for the numerical integration of 
(\ref{NCSE}), we need the small-$x$ behavior of $\gamma_1(x)$. A small-$x$ 
analysis of (\ref{NCSE}) gives $\gamma_1 = b\lambda x^4/16d(d+2) + \cdots$, 
where the results in (\ref{SMLC}) were used to show that the last two 
terms in (\ref{NCSE}) are subdominant as $x \to 0$. The required initial
conditions are therefore $\gamma_1(0) = \gamma_1'(0) = 0$. 
As $x \rightarrow \infty$, the last two terms in (\ref{NCSE}) can be 
neglected. The two linearly independent solutions of the simplified 
equations have a power-law tail ($\sim x^{-(\omega + 2{\lambda})}$) and 
a Gaussian tail ($\sim x^q\exp(-{x^2}/8$)) for large $x$, where 
$q=2\lambda-d+\omega$ if $\omega>2$ and $q=2\lambda-d+2$ otherwise. Having 
already found $\lambda$, $\omega$ is chosen on physical grounds in the 
same way as $\lambda$, namely that the coefficient of the power-law term 
in the large-$x$ solution should vanish.  The values of $\omega$ obtained 
are given in Table 2 for $2 \le n \le 5$ and  $1 \le d \le 3$. Note that  
$\omega \to 4$ for $d \rightarrow \infty$, as the OJK result (and its 
generalization to vector fields) is recovered in this limit.

After solving  (\ref{NSE}) and (\ref{NCSE}) for  $\gamma_{0}(x)$ 
and $\gamma_{1}(x)$ we use these results to get the scaling function 
$f_{0}(x)$ and the correction-to-scaling function $f_{1}(x)$ from 
Eqs.\ (\ref{S}) and (\ref{CS}). Figure 1 
shows the scaling functions $f_{0}(x)$ and the correction-to-scaling 
functions $f_{1}(x)$ for $n = 2$ and 3 in 3$d$. The amplitude of 
$f_{1}(x)$ is arbitrary. It is determined by the coefficient $b$ 
introduced in (\ref{L}): the value $b = 2$ was used in Figure 1. The 
scaling functions and the correction-to-scaling functions do not show strong 
dependence on $n$ and $d$ for $n \ge 2$. For $n = 1$ and $n \ge 2$ the 
scaling functions are very different, especially in the small$-x$ region. The 
reason for this is the presence of the sharp interfaces in $n = 1$ systems, 
which lead to a finite slope at the origin in $f_0(x)$ \cite{Review} and 
a cubic small-$x$ behaviour in $f_1(x)$ \cite{BRC}. For $n \ge 2$, the 
small-$x$ is quadratic for $f_0$, and quartic for $f_1$, with logarithmic 
corrections for even $n$.

Within Mazenko theory the correction-to-scaling exponent $\omega$ is 
non-trivial and depends on both $n$ and $d$, with $\omega \le 4$ for 
all $n$ and $d$ in nonconserved $O(n)$ models. The upper bound of $4$ 
is obtained when $d \rightarrow  \infty$ (for any value of $n$) or 
$n \rightarrow \infty$ (for any value of $d$).

A noteworthy feature of Figure 1 is that the correction-to-scaling function 
$f_1(x)$ is much larger than $f_0(x)$ at large $x$ (the same feature was 
found for many of the models studied in \cite{BRC}). This means that, in
fitting data, scaling violations at large-$x$ should be given less weight 
in choosing fitting parameters (e.g.\ the scale length $L(t)$) than 
violations at small or intermediate $x$, because corrections to scaling are 
larger there.

\section{The one-dimensional XY model}
An exact solution of this model was first presented by Newman et al\cite{NBMH}. 
The solution yields an `anomalous' growth law, $L \sim t^{1/4}$, for the 
characteristic length exhibited by the pair correlation function, 
compared with the usual $L \sim t^{1/2}$ growth law of nonconserved 
models. Mazenko theory does not predict this growth law, for the simple 
reason that the theory has been built in such way that it might be expected 
to give qualitatively correct results only for systems with topological 
defects (i.e. $n \le d$), since the $n$-component auxiliary field 
${\vec m}({\bf r},t)$ is defined in terms of the underlying defect 
structure. Despite this, the theory does a reasonable job of accounting 
for the behavior of systems with $n>d$, and in fact becomes exact in the 
limit $n \to \infty$. However, systems with $n=d+1$, which can support 
topological textures, are poorly treated by this approach. 

An exact solution for the nonconserved $n=2$, $d=1$ system  is possible 
because the equation of motion for the order parameter becomes linear in 
the angle representation, $\vec\phi = (\cos\theta,\sin\theta)$, which is 
natural in the hard-spin limit, where $\vec\phi^2 = 1$. In this limit 
the free energy functional is simply 
$F = (1/2) \int dx (d\vec\phi/dx)^2 = (1/2)\int dx (d\theta/dx)^2 $. 
The zero-temperature equation of motion for model A,  
$\partial\vec\phi/\partial t = -\delta F/\delta\vec\phi$,  becomes, 
in the angle representation,
\begin{equation}
\frac{\partial\theta}{\partial t} = \frac{\partial^2\theta}{\partial x^2}, 
\label{NDE} 
\end{equation}
which is a diffusion equation for the phase angle $\theta$. Thus one
characteristic length scale is the `phase diffusion length', 
$L_\theta = t^{1/2}$, but this is not the scale which characterises the 
pair correlation function. 

Equation (\ref{NDE}) can be solved in Fourier space to give 
${\theta_{k}}(t) = {\theta_{k}}(0)\exp(-{k^2}t)$. In evaluating quantities 
of interest such as correlation functions, one needs to specify the initial 
conditions. The probability distribution, $P([\theta_k(0)])$, for 
$\theta_k(0)$ is conveniently chosen to be Gaussian 
$P([\theta_k(0)]) \propto \exp 
\left[-\frac{1}{2}\sum_k\beta_k\theta_k(0)\theta_{-k}(0) \right ]$. 
The choice $ \beta_k = \xi_0k^2/2$ is made as it gives the initial condition  
$C(r,0) = \exp(-r/{\xi_{0}})$ for the order-parameter correlation function, 
which is the appropriate form for systems quenched from an equilibrium 
disordered state with correlation length $\xi_{0}$. The equal-time 
correlation function is given by 
\begin{equation}
C(r,t) = \langle {\vec \phi}(x,t) \cdot {\vec \phi}(x+r,t)\rangle  
=\langle \cos[\theta(x+r,t) - \theta(x,t)]\rangle.
\label{NMCFE} 
\end{equation}
Using the Gaussian probability distribution for $\theta_{k}(0)$ equation 
(\ref{NMCFE}), with the dynamics (\ref{NDE}) gives
\begin{equation}
C(r,t) = \exp \left(-\sum_{k} \frac{1}{\beta_{k}}\exp(-2{k^2}t)[1-\cos{kr}] 
\right). 
\label{NMCWE} 
\end{equation}
Since the characteristic value of $k$ in the integral is of order 
$t^{-1/2}$, and we anticipate (see below) the growth law $L(t) \sim t^{1/4}$, 
which sets the characteristic scale of $r$ in $C(r,t)$, it follows 
that the scaling limit and the corrections to it can be obtained 
from a power-series expansion of $\cos(kr)$ in (\ref{NMCWE}), since the 
characteristic value of $kr$ is small (of order $t^{-1/4}$) at late times. 
Retaining the leading and next-to-leading terms in the exponent, and 
evaluating the sums over $k$, gives
\begin{eqnarray}
C(r,t) & = & \exp \left[-\frac{r^2}{2\xi_0 (2\pi t)^{1/2}} +  
\frac{r^4}{96\xi_0 (2\pi)^{1/2}t^{3/2}} 
+ O\left(\frac{r^6}{\xi_0 t^{5/2}} \right) \right] \nonumber \\
   & = &\exp \left[-\frac{y^2}{2(2\pi)^{1/2}} \right]\,\left[1 +
\frac{\xi_0^2}{L^2}\,\frac{y^4}{96(2\pi)^{1/2}} + 
O\left(\frac{\xi_0^4}{L^4}\right) \right],   
\label{NMCE} 
\end{eqnarray}
where $y = r/L$ is the scaling variable and the coarsening scale 
$L(t) = \xi_0^{1/2}t^{1/4}$. 
The correction-to-scaling exponent is $\omega = 2$.

This growth law is rather unusual  since the generic form of the growth law 
for nonconserved fields is $L(t) \sim t^{1/2}$. In this model $\omega$ is 
found to be trivial while within Mazenko theory $\omega$ is non-trivial.  
There are two fundamental length scales in this problem, namely the phase 
coherence $t^{1/2}$ and the correlation length $\xi_0$ associated with the 
initial conditions. The coarsening scale $L(t)$ of the pair correlation
function is the geometric mean of these two lengths. 
Note that the pair correlation function has a strong dependence on $\xi_0$, 
which is not `forgotten' at late times. This sensitivity to initial 
conditions is absent in other models, such as  
$ n = \infty$ vector model, where the initial conditions drop out at late 
times. In the conserved 1-$d$ XY model also, simulation 
results gives $L \approx t^{1/6}$ \cite{MG} instead of the  
$L \sim t^{1/4}$ behavior expected in higher dimensions \cite{BR}, 
suggesting that this `anomalous' behaviour may be present 
there also. Although no exact solution is known for the conserved case, 
heuristic arguments, based on the role of the two characteristic lengths, 
can account for the observed $t^{1/6}$ growth \cite{RB}.

\section{Conserved O(n) model}
The dynamical scaling properties of systems with a conserved order 
parameter (Model B) with $O(n)$ symmetry is studied using Mazenko theory. 
Naive application of this theory does not give correct growth 
law $L \sim t^{1/3}$ for scalar fields (the bulk diffusion field must be 
included in order to get the correct law\cite{Maz}). Here, however, we 
will only consider systems with $n \ge 2$. For Model B systems, equation 
(\ref{ACFE}) becomes
\begin{equation}
\frac{1}{2}\,\frac {\partial C(12)}{\partial t} =
-\nabla^2\left[\nabla^2 C(12)+
 \alpha (t)\,\gamma\,\frac{dC(12)}{d\gamma}\right]\ ,  
\label{CCFE}
\end{equation}
with $\alpha(t) = 1/2S_0$. For eq. (\ref{CCFE}) to have a scaling solution  
it is clear that $\alpha\sim 1/L^2$ and $L \sim t^{1/4}$. The latter 
is the correct growth law for $n \ge 2$, but for $n = 2$, $d > 2$ there are 
logarithmic corrections\cite{BR} which (\ref{CCFE}) fails to predict. 
We will first consider the case where $n$ is very large. In this case an 
expansion in $1/n$ can be made in equation (\ref{CCFE}). For large $n$, 
$C(\gamma)\sim\gamma-\gamma(1-\gamma^2)/2n + O(1/n^2)$ and $\gamma\,
\frac{dC(12)}{d\gamma} = C + C^3/n + O(1/n^2)$. With the above truncations, 
(\ref{CCFE}) can be written as
\begin{equation}
\frac{1}{2}\,\frac {\partial C(12)}{\partial t} =
-\nabla^2\left[\nabla^2 C(12)+
 \alpha(t)\,\left(C + \frac{C^3}{n} \right) \right]\,,  
\label{LNSE}
\end{equation}
correct to order $1/n$.

It is worth mentioning that the $C^3/n$ term is essential for scaling 
to be recovered at finite $n$. For $n$ strictly infinite the $C^3/n$ term 
is absent and `multiscaling' is obtained \cite{CZ}. For arbitrary $n$, an 
expansion in powers of $C$ can be made. Truncating the expansion at order 
$C^3$ leads back to (\ref{LNSE}) but with $n$ replaced by an effective 
$n^*$, given by $n^* = (n+2)\/a_n^2$ with 
$a_n = n[B((n+1)/2,1/2)]^2/2\,\pi$ \cite{FA}. 

Dimensional analysis of (\ref{LNSE}) requires $\alpha(t) = \alpha/L^2$, 
which defines $L$. Including the leading corrections to scaling as usual 
we write 
\begin{eqnarray}
C(r,t) & = & f_0(r/L) + L^{-\omega} f_1(r/L) + \cdots \\
dL/dt & = & 1/4{L^3} + b/L^{\omega + 3} + \cdots\ ,   
\end{eqnarray}
where $b$ fixes the amplitude of $f_1(r/L)$. Inserting these expansions 
into (\ref{LNSE}) and comparing terms of leading order, $O(1/L^4)$, and 
next-to-leading order, $O(1/L^{(4 + \omega)})$, gives 
\begin{eqnarray}
\label{LNSF}
\frac{x}{8}\,\frac {df_0}{dx} & = & 
\nabla_x^2 \left[\nabla_x^2 f_0 + 
\alpha\,\left(f_0 + \frac{f_0^3}{n} \right) \right] \\  
\frac{x}{8}\,\frac {df_1}{dx} + \frac{\omega}{8}\,f_1 
+ \frac{bx}{2}\,\frac{df_0}{dx} & = & 
\nabla_x^2\left[\nabla_x^2 f_1 +
 \alpha\,\left(f_1 + \frac{3 f_0^2 f_1}{n} \right) \right]\ ,  
\label{LNCF}
\end{eqnarray}
where $\nabla_x^2 = d^2/dx^2 + [(d - 1)/x]\,d/dx$. 

For general $n$ one must solve (\ref{CCFE}) with $C(r,t)$ given by 
(\ref{BPT}). However, the singularities of $C(\gamma)$ and its 
derivatives at $\gamma = 1$ introduce some numerical difficulties. 
Instead, therefore, we solve (\ref{LNSE}) which is valid for large $n$. 
For general $n$, an expansion in $C$ up to $C^3$, leading to (\ref{LNSE}) 
with an effective $n^*$ \cite{FA}, gives scaling functions 
which are in fairly good agreement with simulation results \cite{FA,SR}.  

In solving (\ref{LNSF}) numerically, one must know the boundary 
conditions. These are provided by small-$x$ and large-$x$ analyses. 
For small-$x$, the series expansion $f_0 = 1 + 
\sum_{r=1}^\infty \beta_r x^r$, substituted into (\ref{LNSF}), 
gives $f_0 = 1 + \beta x^2 - (1+3/n)\,\alpha\,\beta\,
x^4/{4(2+d)} + \cdots$, with $\beta_{2} = \beta$. Numerical integration 
can therefore be performed on (\ref{LNSF}) with initial conditions $f_0(0) 
= 1$, $f_0''(0) = 2\beta$, $f_0'(0) = f_0'''(0) = 0$. Both 
$\alpha$ and $\beta$ are undetermined parameters. 

For the large-$x$ analysis, we impose the physical condition that 
$f_0(x) \rightarrow 0$ for $x \to \infty$. This leads to the linearised 
version of eq. (\ref{LNSF}) given by
\begin{equation}
\frac{x}{8}\,\frac {df_0}{dx} = \nabla_x^2 \left[\nabla_x^2 f_0 +
\alpha\,f_0  \right]\,.  
\label{LIVE}
\end{equation}
There are four linearly independent solutions of  (\ref{LIVE}), with the 
general asymptotic form  
\begin{equation}
f_0(x) \sim F_0 x^c \exp(-Bx^v-Ax^s)\ .
\label{asymp}
\end{equation} 
The first solution is the constant solution, corresponding to $A = c = B = 0$. 
It satisfies (\ref{LIVE}) by inspection. The other three solutions are 
obtained by  substituting (\ref{asymp}) into (\ref{LIVE}) 
and carrying out an asymptotic large-$x$ analysis, leading to the relations 
\begin{eqnarray}
v & = & 4/3\,, \nonumber\\
s & = & 2/3 \,,\nonumber\\
B^3 & = & -1/8v^3\,, \nonumber\\
A & = & 64\alpha B^2/9\,,  \nonumber\\
c & = & -2d/3\,. \nonumber
\end{eqnarray}
The three different solutions correspond to the three solutions for $B$, 
one real, two complex. The real solution, $B = -1/2v = -3/8$, leads to 
an  exponentially diverging solution for $f_0$:  
\begin{equation}
f_0(x) \sim F_0\,x^{-2d/3}\exp(3x^{4/3}/8 - \alpha x^{2/3}),
\end{equation}
while the two complex roots, $B = 3(1 \pm i\sqrt{3})/16$, generate 
two solutions which can be combined to give an exponentially decaying 
solution with oscillatory behaviour
\begin{equation}
f_0(x) \sim F_0 x^{-2d/3}\exp\left(-\frac{3x^{4/3}}{16} + \frac{\alpha 
x^{2/3}
}{2}\right)\cos\left(\frac{3\sqrt{3}x^{4/3}}{16}+ 
\frac{\alpha\sqrt{3}x^{2/3}}{2} + \varphi_0\right)\,,
\end{equation}
where $F_0$ and $\varphi_0$ are arbitrary constants. 

Just as in the Model A case, where $\lambda$ was fixed by imposing 
physical conditions on the large-$x$ solution, also in this case we 
have an eigenvalue problem in which two parameters $\alpha$ and $\beta$ are 
chosen to eliminate the unphysical constant solution and the exponentially 
diverging solution. The same problem is encountered in Model B with a 
scalar order parameter \cite{MAZ2}. Applying the procedure described in  
\cite{FA,MAZ2} it is possible to determine $\alpha$ and $\beta$. 

Turning now to the corrections to scaling, we consider first the 
four linearly independent large-$x$ solutions for the 
linearised form of eq. (\ref{LNCF}). These are a power law solution, 
$f_1(x) \sim x^{-\omega}$, an exponentially growing solution, $\sim x^p 
\exp(3x^{4/3}/8 - \alpha x^{2/3})$, and two decaying solutions that 
can be combined in the form   
\begin{equation}
f_1(x)\sim x^p \exp\left(\frac{-3x^{4/3}}{16} +
\frac{\alpha x^{2/3}}{2}\right)
\cos\left(\frac{3\sqrt{3}x^{4/3}}{16} 
+  \frac{\alpha \sqrt{3} x^{2/3}}{2} + \varphi_1\right)\,,  
\end{equation}
where $\varphi_1$ is arbitrary, and $p = (\omega - 2d)/3$ if $\omega > 4$ 
and $p = (4 - 2d)/3$ otherwise. The small-$x$ solution 
is $f_{1}(x) = \mu x^2 - \alpha\, \mu\, (1+3/n)\,x^4/4(d+2) + \cdots$ . 
Therefore (\ref{LNCF}) is solved numerically with initial conditions 
$f_1''(0) = 2\mu$, $f_1(0)=f_1'(0)=f_1'''(0)=0$. The two parameters 
$\mu$ and $\omega$ are as yet undetermined. They are fixed in the same 
way as $\alpha$ and $\beta$, by requiring that an oscillatory, 
exponentially decaying solution is recovered as $x\rightarrow\infty$.

Values for $\alpha$, $\beta$, $\mu$ and $\omega$ in 3-$d$ for $n = 2$, 5, 
10, 20 and  50 are shown in Table III (For $n = 2$, the effective 
$n^* = \pi^2 /4$ has been used). The functions $f_0(x)$ and  $f_1(x)$ are 
displayed in Figure 2 for $n = 2$ and 20 in 3-$d$. Again $b$ has been set 
to $b = 2$ without loss of generality. 

The most important result to be extracted for Table III is that the value 
of $\omega$ decreases as $n$ increases. This behaviour is quite different 
from Model A, where $\omega$ increases asymptotically to $4$ as $n$ 
increases. It seems from Table III that $\omega$ probably tends to zero for 
$n \to \infty$, although an analytical determination of the correction to 
scaling for $n$ large but finite, analogous to the treatment of the leading  
scaling function in \cite{BH2}, has not yet been realized.  

Comparison of the scaling and correction-to-scaling functions displayed in 
Figure 2 reinforces a point made in connection with the nonconserved systems, 
namely that the correction to scaling become large (relative to the scaling 
function itself) at large values of the scaling variable $x$. As noted 
before, this suggests that in carrying out scaling analyses of data, 
more attention should be paid to small and intermediate values of $x$, where 
corrections to scaling can be expected to be (relatively) smaller, than to 
large $x$. Indeed, for the nonconserved case (Figure 1) the correction to 
scaling has its maximum at a point where the scaling function is already 
quite small (around 0.1).

\section{Summary}
Corrections to scaling associated with a non-scaling initial condition have 
been studied in $O(n)$ models within the Gaussian closure scheme of Mazenko. 
We have calculated both the correction-to-scaling function, $f_1(x)$, and the 
associated correction-to-scaling exponent, $\omega$, for both nonconserved and 
conserved fields. In both cases Mazenko theory suggests that $\omega$ is 
nontrivial, depending on the nature of the dynamics involved, the 
dimensionality, $d$, of the system and the number of order parameter 
components, $n$. For nonconserved fields the value of  $\omega$ tends 
to the limiting value $4$ for $n \to \infty$ with $d$ fixed, and for  
$d \to \infty$ with $n$ fixed. In the latter limit, the Mazenko theory 
reduces \cite{BH1,LiuMaz} to the OJK theory \cite{OJK} and its 
generalizations \cite{BPTO}, believed to become exact as 
$d \to \infty$ \cite{LIME}.

The 1-$d$ XY model is anomalous in that it exhibits a different growth law 
from the standard one for nonconserved dynamics, and the correction-to-scaling 
exponent is simple ($\omega = 2$). In this model quantities of interest, 
such as the correlation function $C(r,t)$,  retain `memory' of the initial 
conditions even in the scaling limit.

In studying the conserved $O(n)$ model, an expansion in $1/n$ was used which 
is valid for large $n$. This approach was used to find the 
correction-to-scaling function $f_1(x)$ and the exponent $\omega$ for 
$n = 5$, 20 and 50 in 3-$d$. For $n=2$, an expansion in $C$ up to 
$C^3$ was made. In the latter case, a comparison \cite{FA} between 
leading-order scaling results and simulations shows very good agreement 
despite the wrong growth law (i.e without the logarithmic corrections 
predicted for $n=2$ \cite{BR}). In conserved systems $\omega$ decreases 
as $n$ increases, raising the question of whether $\omega \rightarrow 0$ or 
approaches some limiting value as $n$ becomes very large. We have as yet 
been unable to find $f_1(x)$ and $\omega$ analytically in the limit of 
large but finite $n$ -- this remains an interesting open question. 

The main lesson for the analysis of experimental and simulation data is 
that corrections to scaling can be expected to be relatively small at small 
and intermediate scaling variable $x$ (=$r/L$), suggesting that this region 
be given more weight than large $x$ in fitting (or collapsing) data.   

\bigskip

\begin{center}
\begin{small}
{\bf ACKNOWLEDGEMENTS}
\end{small}
\end{center}
This work was supported by EPSRC under grant GR/L97698 (AB) and 
by the Commonwealth Scholarship Commission (NR).

\newpage 

\begin{table}
\begin{center}
\caption{Exponent $\lambda$ within Mazenko Theory for model A.}
\vspace{0.3in}
\begin{tabular}{|c|c|c|c|c|}
\hline
$\hspace{0.1in}n\hspace{0.1in}$& 2 & 3 & 4 & 5 \\ 
\hline
$\hspace{0.1in}d = 1\hspace{0.1in}$&0.301&0.378&0.414&0.433\\ 
\hline
$\hspace{0.1in}d = 2\hspace{0.1in}$&0.829&0.883&0.912&0.930 \\ 
\hline
$\hspace{0.1in}d = 3\hspace{0.1in}$&1.382&1.413&1.432&1.445 \\    
\hline                            
\end{tabular}
\end{center}
\end{table}

\begin{table}
\begin{center}
\caption{Correction-to-scaling exponent $\omega$ within Mazenko Theory 
for model A.}
\vspace{0.3in}
\begin{tabular}{|c|c|c|c|c|}
\hline
$ n $ & 2 & 3 & 4 & 5\\ 
\hline
$d = 1$ & 3.976 & 3.982 & 3.990 & 3.993   \\ 
\hline
$d = 2$ & 3.928 & 3.946 & 3.961 & 3.970  \\ 
\hline
$d = 3$ & 3.930 & 3.945 & 3.958 & 3.966  \\    
\hline                            
\end{tabular}
\end{center}
\end{table}

\begin{table}
\begin{center}
\caption{Values for the eigenvalues $\alpha$, $\beta$, $\mu$ and  
$\omega$ within Mazenko Theory for model B.}
\vspace{0.3in}
\begin{tabular}{|c|c|c|c|c|}
\hline
$ n $ & $\alpha$ & $\beta$ & $\mu$ & $\omega$         \\ 
\hline
2     & 1.54880435  & -0.3336250 & 0.227495 & 2.4613967  \\ 
\hline
5     & 1.72743447  & -0.3179775 & 0.225025 & 2.0667992  \\ 
\hline
20    & 2.01748270  & -0.3292250 & 0.214515 & 1.1290901  \\    
\hline                            
50    & 2.179330049 & -0.3487125 & 0.206515 & 0.5029987  \\ 
\hline                            
\end{tabular}
\end{center}
\end{table}

\newpage

\begin{center}
FIGURES
\end{center}

\underline{Figure 1.} The scaling function $f_{0}(x)$ and the 
correction-to-scaling function $f_{1}(x)$ for nonconserved order parameter.  
Continuous and broken lines correspond to $d = n = 3$ and $d = 3$, $n = 2$  
respectively. 

\underline{Figure 2.} Same as in Figure 1, but for conserved order 
parameter. Continuous and broken lines correspond to $d=3$, $n = 20$ and 
$d=3$, $n=2$  respectively (note the different scales for the $x$-axes 
in the upper and lower plots). 

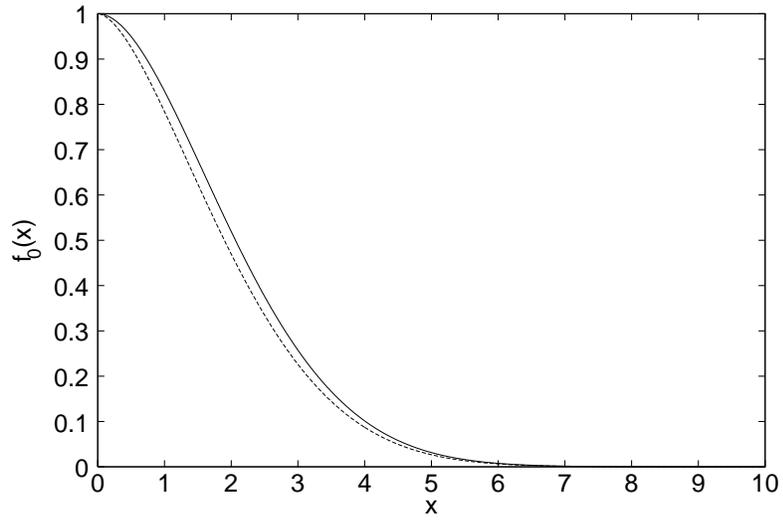
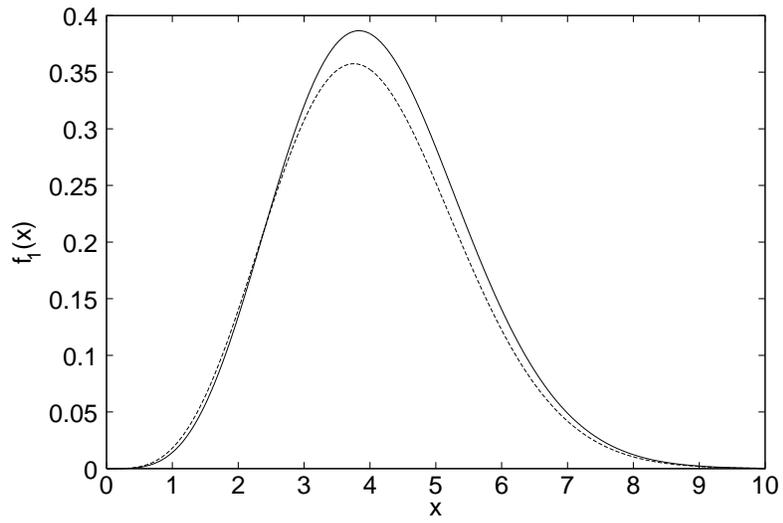
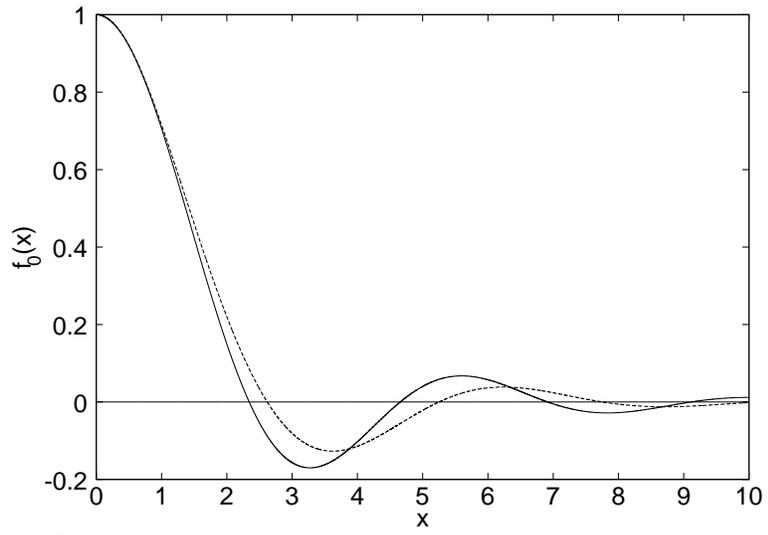
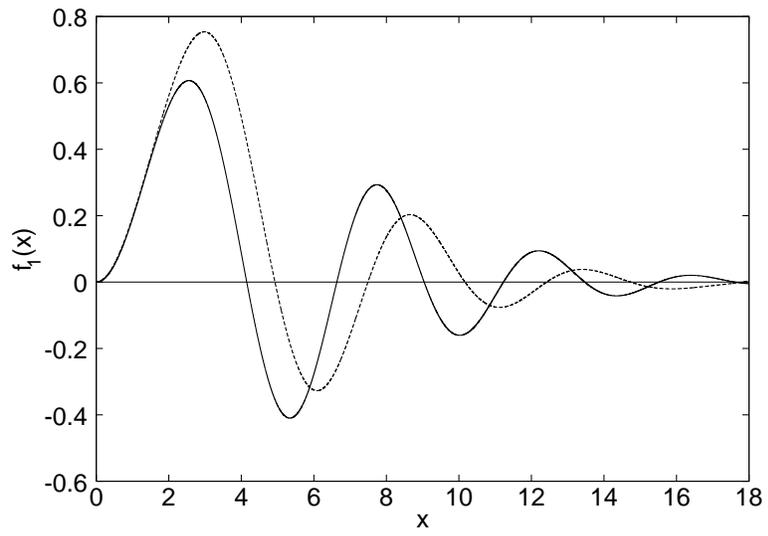
\end{document}